%% file: paper.tex
\documentclass{article}
\usepackage{spconf,amsmath,graphicx}

\usepackage{hyperref}

\title{Detecting Speech Abnormalities with a Perceiver-based Sequence Classifier that leverages a Universal Speech Model}

\name{
\begin{tabular}{cc}
Hagen Soltau$^1$, Izhak Shafran$^1$, Alex Ottenwess$^2$, Joseph R. JR Duffy$^3$, Rene L. Utianski$^3$ \\ Leland R. Barnard$^3$, John L. Stricker$^3$, Daniela Wiepert$^3$, David T. Jones$^3$, Hugo Botha$^3$
\end{tabular}
}
\address{$^1$Google DeepMind, $^2$ Google, $^3$Mayo Clinic}

\copyrightnotice{979-8-3503-0689-7/23/$31.00 ©2023 IEEE}

\begin{document}
\maketitle
\input{abstract}

\begin{keywords}
speech disorders, neurological tests, sequence classification, perceiver
\end{keywords}

\input{intro}
\input{related_work}
\input{data}
\input{experiments}
\input{conclusions}

\section{Acknowledgements}
We are grateful for help, discussion, and support from Greg Corrado, Anthony Phalen, Anil Chintapatla, Enrique Chan, Megan Freedman, Zoubin Ghahramani, and Natalie Miller.

\bibliographystyle{IEEEbib}
\bibliography{paper}

\end{document}

%% file: abstract.tex
\begin{abstract}
We propose a Perceiver-based sequence classifier to detect abnormalities in speech reflective of several neurological disorders. We combine this classifier with a Universal Speech Model (USM) that is trained on 12 million hours of diverse audio recordings. Our model compresses long sequences into a small set of class-specific latent representations and a factorized projection is used to predict different attributes of the disordered input speech. The benefit of our approach is that it allows us to model different regions of the input for different classes and is at the same time data efficient. We evaluated the proposed model extensively on a curated corpus from the Mayo Clinic. Our model outperforms standard transformer (80.9\%) and perceiver (81.8\%) models and achieves an average accuracy of 83.1\%. With limited task-specific data, we find that pretraining is important and surprisingly pretraining with the un-related automatic speech recognition (ASR) task is also beneficial. Encodings from the middle layers provide a mix of both acoustic and phonetic information and achieve best prediction results compared to just using the final layer encodings (83.1\% vs 79.6\%). The results are promising and with further refinements may help clinicians detect speech abnormalities without needing access to highly specialized speech-language pathologists.
\end{abstract}

%% file: intro.tex
\section{Introduction}
\label{sec:intro}

Neurological injury or illness is one of the leading causes of mortality and morbidity globally~\cite{feigin2017global}. A key contributor to this is the lack of access to expertise, leading to diagnostic delays, misdiagnosis, and lack of appropriate treatment~\cite{janca2006wfn}. Many neurological conditions manifest in speech. For example, slow or breathy speech defines motor speech disorders (e.g., the dysarthrias) which has specific associations with neurological diseases (e.g., ALS).

Speech and language pathologists (SLPs) administer clinical evaluations to uncover abnormalities in specific speech characteristics to aid differential diagnosis. Unfortunately, this expertise is not widely available. An automated method to detect these characteristics would have the potential to improve diagnostic accuracy and treatment for neurological disease~\cite{dang2021progressive}.

In this paper, we develop models to detect abnormalities in speech across three commonly used tasks: vowel prolongation (VP), alternate motion rate (AMR), and sequential motion rate (SMR).
We do so in a dataset that is large from a clinical speech corpus standpoint, but very small from a deep learning one.
Labelled data in this domain is difficult to obtain because of concerns around sharing health data and labeling the data carefully is labor intensive and requires domain expertise that is rare.

We overcome the data limitation by leveraging foundation models which are trained on large amounts of data. While there has been a long history of developing ``universal'' models in speech where the definition of ``universal'' has changed over time -- robustness to a variety of acoustic environments, domain independence and multi-linguality, the current versions of foundational models are based on orders of magnitude mode data and parameters~\cite{zhang2023google}. These foundation models can generate powerful representations of speech, modeling long temporal dependencies (up to 120 sec in our corpus). However, extracting relevant signals for sequence classification remains a challenging problem. In this work, we propose a perceiver~\cite{jaegle2021perceiver} based architecture for sequence classification. The model maps a variable-length sequence to a small, fixed, set of class-specific latent representations, making it very suitable for sequence classification tasks.

To place this work in the context of literature, we discuss related work in Section~\ref{sec:related_work}. In Section~\ref{sec:data}, we describe three clinical tasks on which we evaluate our proposed model. Our perceiver-based model is described in Section~\ref{sec:experiments}, where we also report experimental results and discuss them. The key takeaways from this work is summarized in Section~\ref{sec:conclusions}.

%% file: related_work.tex
\section{Related Work}
\label{sec:related_work}
There have been numerous prior studies that have used machine learning to detect or quantify speech or neurological disorders using short speech recordings. 
For example, feature extraction followed by traditional statistical or classical machine learning models have been applied to Amyotrophic Lateral Sclerosis (ALS)~\cite{vashkevich_classification_2021, tena_detecting_2022}, Alzheimer's disease (AD)~\cite{hajjar_development_2023}, Parkinson's disease (PD)~\cite{singh_robust_2020,parkinsons2015}, depression~\cite{depression2014},
and autism~\cite{rybner_vocal_2022}.

Even though deep learning on raw or unstructured audio outperform traditional machine learning approaches on most speech-based tasks, such as emotion recognition~\cite{ragheb_emotional_2022}, few prior studies have applied them to neurological or speech disorders. A recent paper used Mel-spectrograms and convolutional deep networks to separate Parkinson's disease from healthy controls using sustained vowel (accuracy 81.6\%) or a combination of sounds (92\%)~\cite{quan_end--end_2022}. 
Another paper used a auto-encoder to learn representations of the spectrograms and then trained a classifier on top of these to separate healthy controls from those with mild cognitive impairment or dementia with an accuracy of 90.57\%~\cite{bertini_automatic_2021}. Use of pretrained models is even rarer. One recent paper used a pretrained data2vec~\cite{baevski_data2vec_2022} model to extract voice embeddings which were then used in a classifier to detect AD~\cite{agbavor_artificial_2023}.

At face value the accuracies reported in these studies are encouraging, but unfortunately common methodological weaknesses limit their real world applicability. For example, the binary classification into healthy vs one disease entity is not reflective of the complex task healthcare providers face, who may be able to recognize the speech is abnormal but lack the expertise to determine which specific disease is the likely cause. Moreover, many studies do not use a true test or hold out set, and most have very small sample sizes. The commonly used UA-Speech and TORGO corpus for dysarthria have 15 and 7 speakers with speech disorders, respectively~\cite{kim_dysarthric_nodate,rudzicz_torgo_2012}. These small sample sizes raise validity concerns, and recent work suggests that models trained on the non-voiced portions of these corpora may outperform those that include the voiced sections~\cite{schu_using_2022}. The corpora used are often highly curated, and the performance on these sets do not translate to more realistic corpora in neurological disease~\cite{arora_developing_2019}.

From a modeling point of view, the task is essentially a sequence classification task. Text-based sequence classifiers often make use of a 'classifier token' that summarizes the sentence~\cite{bert, radford2018improving}. While this approach could potentially be applied in the audio domain, it would be fairly sub-optimal for our task, as different classes need cues from different regions of the signal. Alternatively, I-vectors~\cite{ivector} are used to condense audio sequences to a fixed length representation for example in speaker verification. LSTMs provide a 'neural' version where the state vector can provide a fixed length representation of the input sequence~\cite{Hochreiter1997}. But both of them share the same limitation that the predicted classes cannot 'look' at different regions of the signal.

Attention with latent units was introduced in the Set Transformer model ({\em induced} set attention block)~\cite{lee2019set} as a way to reduce the quadratic time complexity of attention functions. Note, in their work the output dimension remains the same as the input dimension. In our case, we are interested to map variable length sequences to a small number of fixed length representations, but otherwise rely on the same mechanism. In the perceiver model~\cite{jaegle2021perceiver}, cross-attention with a set of latent units is applied in a iterative fashion to compress high-dimensional different modalities and integrates them with transformer blocks. The logits of their model are formed by averaging the outputs of the self-attention block, similar to classification with transformers. Note, this is different in our model, where no pooling operation needs to be performed, instead we rely on a factorized projection, see section~\ref{sec:perceiver_model}. In~\cite{wolters2021proposalbased}, a region proposal network is combined with a perceiver module for sound event detection, however the details are missing and the experimental setup relies on artificially created examples.





%% file: data.tex
\section{Task description}
\label{sec:data}

The goal of this project is to aid clinicians in detecting abnormal speech characteristics, which may be helpful in diagnosing neurological conditions. As mentioned in section~\ref{sec:related_work}, existing studies face limitations in their experimental setup. We address these limitations by using a much larger dataset of recordings collected in a clinical setting with a goal of detecting several abnormal speech characteristics across a range of tasks. In this study, we focus on the following three tasks:
\begin{enumerate}
  \setlength{\itemsep}{0pt}
  \setlength{\parskip}{0pt}
    \item Vowel Prolongation: Patients are asked to take a deep breath and make an ‘aaaah’ sound for as long and steadily as they can until they run out of breath. The task is used for isolating respiratory/phonatory characteristics.
    \item Alternating Motion Rate: Patients are asked to take a deep breath and repeat ‘puh’ sounds as fast and regularly as possible. The task is used for assessing speed and regularity of rapid, repetitive movements and helpful in eliciting articulation breakdowns.
    \item Sequential Motion Rate: Patients are asked to take a deep breath and repeat ‘puh-tuh-kuh’ sounds as fast and regularly as possible. The task is used for assessing the ability to produce sound sequences quickly and accurately, helpful for articulatory and prosodic breakdowns.
\end{enumerate}

\subsection{Modeled Attributes (Labels)}

For each task, a set of characteristics (labels) need to be predicted ~\cite{duffy2019motor}. There are $14$ labels in all, and they are listed below:
\begin{enumerate}
  \setlength{\itemsep}{0pt}
  \setlength{\parskip}{0pt}
    \item abnormal loudness variability: can reflect poor coordination or control of respiratory flow
    \item abnormal pitch variability: can reflect poor coordination or control of laryngeal movements
    \item breathy: can reflect vocal fold weakness/ incomplete glottal adduction
    \item distortions: can reflect deficits of programming or weakness
    \item flutter: can reflect vocal fold weakness or rapid tremor
    \item hoarse / harsh: can reflect excess glottal adduction
    \item irregular articulatory breakdowns: can reflect poor coordination
    \item loudness decay: can reflect reduced respiratory or laryngeal function
    \item rapid rate: can reflect deficit in basal ganglia control circuit
    \item slow rate: can reflect deficits of programming, control, or execution
    \item strained: can reflect vocal fold hyperadduction
    \item syllable segmentation: can reflect poor programming or coordination
    \item tremor: can reflect deficit in basal ganglia control circuit
    \item unsteady: can reflect poor coordination or control of respiratory and/or laryngeal movements
\end{enumerate}

Binary ratings for these characteristics were annotated for each recording and task separately, i.e., if a patient had multiple recordings on a specific task they were annotated separately. Since the speech tasks may contain sensitive patient data, the data set cannot be made publicly available. However, to give a sense of this type of data with annotations, we refer to a very small, publicly available vowel prolongation dataset~\cite{dimauro2016voxtester,dimauro2017assessment} of patients with Parkinson's disease that is available at ~\url{https://github.com/Neurology-AI-Program/public\_speech\_data}.

\subsection{Performance metrics}
The overall goal is to predict the presence of every attribute or label for each audio recording. Multiple attributes can be present for a given audio recording. The problem setup is essentially a binary classification task for each of the $14$ attributes. The reported performance metric is the average prediction accuracy.

\subsection{Data and Train/Test partition}
The data set for all three tasks comprises $2594$ audio files with an average length of $37.5$ secs. There are in total $782$ unique speakers, with an average of $3.3$ recordings per speaker. The dataset was roughly balanced by sex. We partitioned the data set into $80\%$ training / $20\%$ test set with {\em non-overlapping} speakers. The split into training vs test data was done randomly in a deterministic fashion. We validated that the experimental setup is insensitive to the chosen partition, e.g. different data splits yield to models with very similar prediction performance.

With the exception of the ablation experiment in section~\ref{sec:task_pooling}, in all experiments the data of all tasks are pooled and the model jointly performs all tasks.

\begin{table}[ht] \centering
   \begin{tabular}{|c|c|c|c|c|c|} \hline
    81.5\% & 81.8\% & 81.6\% &  82.1\% & 81.7\% & 81.7\% \\ \hline
   \end{tabular}
   \caption{Prediction accuracy when training models with different random seed for data partitioning. Experimental setup is largely invariant against different splits of training and test data.} 
   \label{tab:data-split.}
\end{table}

%% file: experiments.tex
\section{Experiments}
\label{sec:experiments}

\begin{figure*}
 \centering 
    \includegraphics[width=0.7\textwidth]{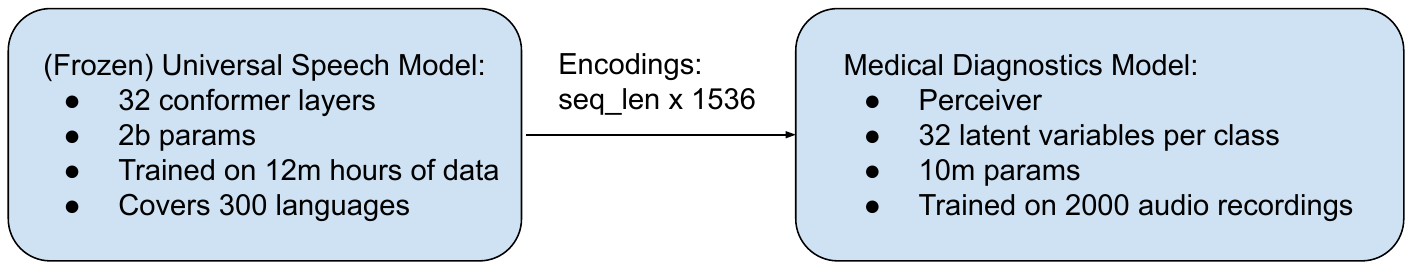}
 \caption{General model architecture. We use a frozen audio encoder, trained on a large and diverse set of speech data. The down-stream model uses cross-attention (Perceiver) to map the encoded audio sequence to a small set of latent representations.}
 \label{fig:model_architecture}
\end{figure*}

The general model architecture is depicted in figure~\ref{fig:model_architecture}.
Since the amount of task specific data is fairly limited, we use a frozen audio encoder (Universal Speech Model) to generate speech representations. The Universal Speech Model is described further in~\cite{zhang2023google}. USM is trained on a large collection of YouTube audio recordings, covering 12m hours of data. The model consists of $32$ conformer blocks~\cite{gulati2020conformer}. The conformer blocks are configured to emit embeddings of $1536$ size as indicated in figure~\ref{fig:model_architecture}.

\begin{figure*}
 \centering 
    \includegraphics[width=0.60\textwidth]{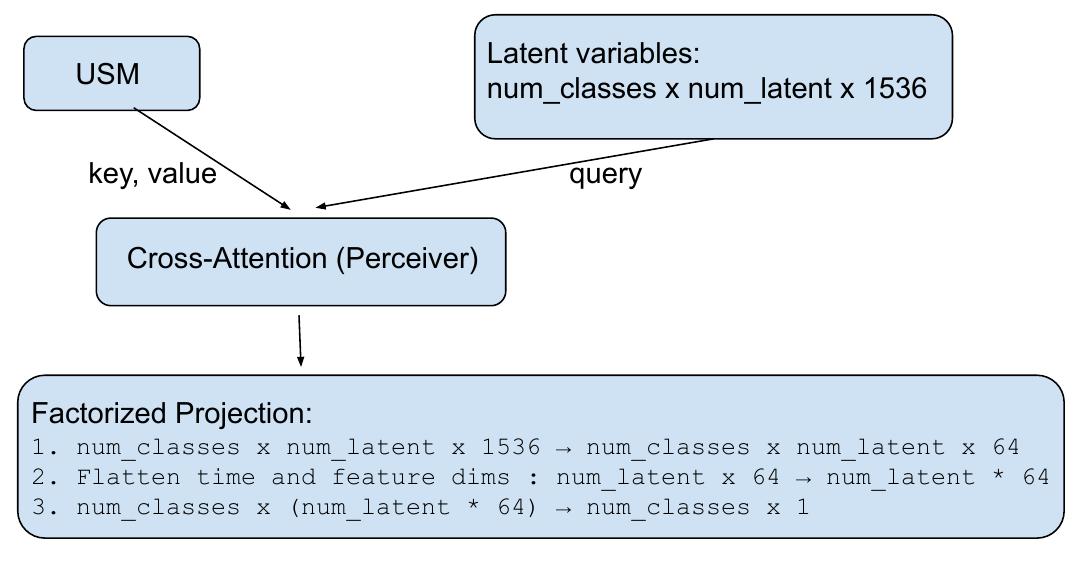}
 \caption{Sequence classification with class specific latent representations.
 In the factorized projection step, the flattening operation collapses time and feature into a single dimension. In the third step, a linear layer operating both in time and feature domain captures joint time-feature dependencies.}
 \label{fig:perceiver}
\end{figure*}

USM is trained in two stages: In the first stage, the model is trained unsupervised using BEST-RQ~\cite{chiu2022selfsupervised}, where a randomly initialized and frozen vector quantizer is used to generate targets so the model can be trained with a reconstruction loss. In the second stage, USM is trained as an ASR model using CTC~\cite{graves_ctc}.

\subsection{Perceiver based architecture}
\label{sec:perceiver_model}
As described in section~\ref{sec:data}, the goal is to model attributes that exhibit temporal structures, e.g. the model needs to be able to capture temporal changes in the signal (e.g., loudness decay, pitch variability). The predominant modeling choice for such problems are transformers~\cite{Vaswani2017}, which we use as one of our baselines.
Following standard practices for classification tasks, the vector sequence from the transformer block is averaged by a pooling operation and logits are computed using a dense layer.

Condensing the sequential information into a pooled vector may not provide enough representational capacity to capture temporal dynamics in long sequences. Perceivers~\cite{jaegle2021perceiver} have been proposed to compress long sequences via cross-attention with learned latent variables and we therefore use perceivers as an additional baseline.

Analogous to perceiver, our model is based on cross-attention with latent variables. The model is illustrated in figure~\ref{fig:perceiver}. Cross-attention maps the encoded audio sequence from USM to a fixed set of {\em class-specific} latent representations. The motivation for this is that the attributes that we want to model are highly diverse and each target class may need to 'look' at different regions of the audio signal. Note, the model parameters of the cross-attention mechanism (key, value, query) are shared across the classes, only the latent variables are class specific - making the model data efficient while allowing for flexibility in temporal modeling.

The logits are calculated using a factorized projection, where we first reduce the feature dimensionality and then use a joint projection operating both in time (latent) and feature axis to produce the final outputs. The  model parameters of the factorized projection are class independent, while the input activations are class specific. 

Compared to Perceiver~\cite{jaegle2021perceiver}, the two key differences are that the latent variables are class specific, which enables modeled classes to use different regions of the signal, and the factorized projection that operates both in time (latent) and feature space allowing for modeling joint dependencies. In contrast, for Perceivers, one common latent representation is used for all classes, and the latent axis is collapsed via average pooling prior of emitting logits.




\begin{table} \centering
   \begin{tabular}{|l|c|} \hline
    Model & Prediction Accuracy \\ \hline
    Transformer & 80.9\% \\ \hline
    Perceiver & 81.8\% \\ \hline \hline
    Our model & 83.1\% \\ \hline
   \end{tabular}
   \caption{Comparison of temporal modeling approaches for sequence classification of medical diagnostics attributes: Our model uses class-specific latent variables for cross-attention with factorized projection. It outperforms strong transformer and perceiver baselines for sequence classification.} 
   \label{tab:transformer-vs-perceiver}
\end{table}

We compare our model against the baselines in table~\ref{tab:transformer-vs-perceiver}. We optimized the hyper-parameters for all approaches. Our model with class-specific latents and factorized projection obtains better performance with $83.1\%$ prediction accuracy compared to both of our baselines. 

\subsection{USM Encodings / Pre-training objective}
In the next set of experiments, we investigate the value of audio encodings from the Universal Speech Model. We choose to base our experiments on USM since prior experiments showed that BEST-RQ outperformed other pre-trained audio encoders~\cite{chiu2022selfsupervised}. Since USM is trained in two stages, BEST-RQ first, followed by CTC, we have the choice of either using encodings that are optimized for ASR or reconstruction loss (BEST-RQ). Since certain attributes (labels) are closer to acoustic characteristics (i.e. loudness decay, pitch variability) while others closer to phonetic characteristics (i.e. syllable segmentation, irregular articulatory breakdown), it is unclear which encodings will be better for the medical diagnostics tasks.

Additionally, we benchmark the model with no pre-training at all. In that case, we skip USM, and directly train our diagnostics model using standard 128-dim log-mel features, operating at a 10 ms frame rate. The results are summarized in table~\ref{tab:pretrain}. In all cases, we use our perceiver approach as described in the previous section. Using USM encodings provide a large gain in prediction accuracy over standard spectrum features, since task specific data is fairly limited. This is very much analogous to the use of large language models~\cite{bert} for NLP tasks. When comparing training USM with BEST-RQ vs ASR, surprisingly the ASR-optimized encodings are better for the medical diagnostics task, even though ASR and the prediction of attributes for speech pathology are substantially different tasks. We shed some light into this by doing a layer-wise analysis in the next section.

\begin{table}[ht] \centering
   \begin{tabular}{|l|c|} \hline
    Audio Encodings & Prediction Accuracy \\ \hline
    No pre-train (log-mel) & 76.4\% \\ \hline
    Unsupervised (BEST-RQ) & 82.0\% \\ \hline
    Supervised (ASR)       & 83.1\% \\ \hline
   \end{tabular}
   \caption{Comparison of using different audio encodings. Audio Encodings from USM give substantial gains in prediction accuracy.} 
   \label{tab:pretrain}
\end{table}

\subsection{Layer encodings}
In the next set of experiments, we investigate whether different (conformer) layers of USM learn different representations. For example, one can argue that representations of higher layers might capture more relevant information w.r.t. the objective function, e.g. ASR. In table~\ref{tab:layers}, we report the prediction accuracy for different layers and find a significant performance variation across layers. The middle layers perform far better than either the lower or higher layers.

\begin{table}[ht] \centering
   \begin{tabular}{|l|c|} \hline
    Layer encodings & Prediction Accuracy \\ \hline   
    layer\_idx=0  & 79.6\% \\ \hline
    layer\_idx=2  & 80.5\% \\ \hline
    layer\_idx=11 & 83.1\% \\ \hline
    layer\_idx=31 & 79.3\% \\ \hline    
   \end{tabular}
   \caption{Prediction accuracy with encodings from different USM layers.} 
   \label{tab:layers}
\end{table}

To get more insights, we analyzed the results w.r.t. individual attributes (labels) and looked which attributes perform better at lower vs higher layers, see table~\ref{tab:layerdetails}. The results suggest that encodings from the first layer retain more relevant acoustic signals while the last layer provides more phonetic or semantic information. This analysis suggests our model can be improved further by allowing attributes (labels) to access audio encodings from different layers. We will explore this in future work.

\begin{table}[ht] \centering
   \begin{tabular}{|l|c|c|} \hline
    Attribute (Label) & \multicolumn{2}{|c|}{Prediction Accuracy} \\   
    & layer\_idx=0 & layer\_idx=31 \\ \hline
    syllable segmentation & 83.0\% & {\bf 88.7\%} \\ \hline
    irregular artic. breakdowns & 72.3\% & {\bf 75.9\%} \\ \hline
    loudness decay & {\bf 80.2\%} & 75.1\% \\ \hline
    breathy & {\bf 81.0\%} & 76.4\% \\ \hline
   \end{tabular}
   \caption{Analysis of prediction accuracy with encodings from different conformer layers w.r.t. different attributes (labels)} 
   \label{tab:layerdetails}
\end{table}

\subsection{Task Pooling}
\label{sec:task_pooling}
As described in section~\ref{sec:data}, by 'task' we refer to the {\em task} a patient is performing, e.g. vowel prolongation. Most tasks have considerable overlap in the predicted label space. For example, 'breathy' and 'strained' are labels both for vowel prolongation as well as alternating motion rate. However, the corresponding audio signals represent different speech patterns, e.g. a long 'aaaah' sound vs repeated 'puh' sounds.

Pooling the data across tasks and building a joint model for all tasks would have two benefits: First, model quality might improve since the amount of training data is fairly limited. Secondly, it makes {\em deployment} of the model substantially easier.

We performed an ablation study where we measure the performance of common labels for the vowel prolongation and alternating motion rate task. The results in table~\ref{tab:taskpool} show modest improvements from pooling data from different tasks. However, the main benefit is that a joint model for all tasks makes deployment much easier. For the final model, we added also the task id as an input to the model, which gave additional gains.

\begin{table}[ht] \centering
   \begin{tabular}{|l|c|} \hline
    Training data & Shared labels for Vowel task \\ \hline   
    Vowel task  & 83.6\% \\ \hline
    Vowel + AMR task & 84.1\% \\ \hline
   \end{tabular}
   \caption{Prediction accuracy when pooling data across tasks.} 
   \label{tab:taskpool}
\end{table}

\begin{figure}
 \centering 
    \includegraphics[width=0.5\textwidth, trim= 0.0in 0.0in 0.0in 0.0in]{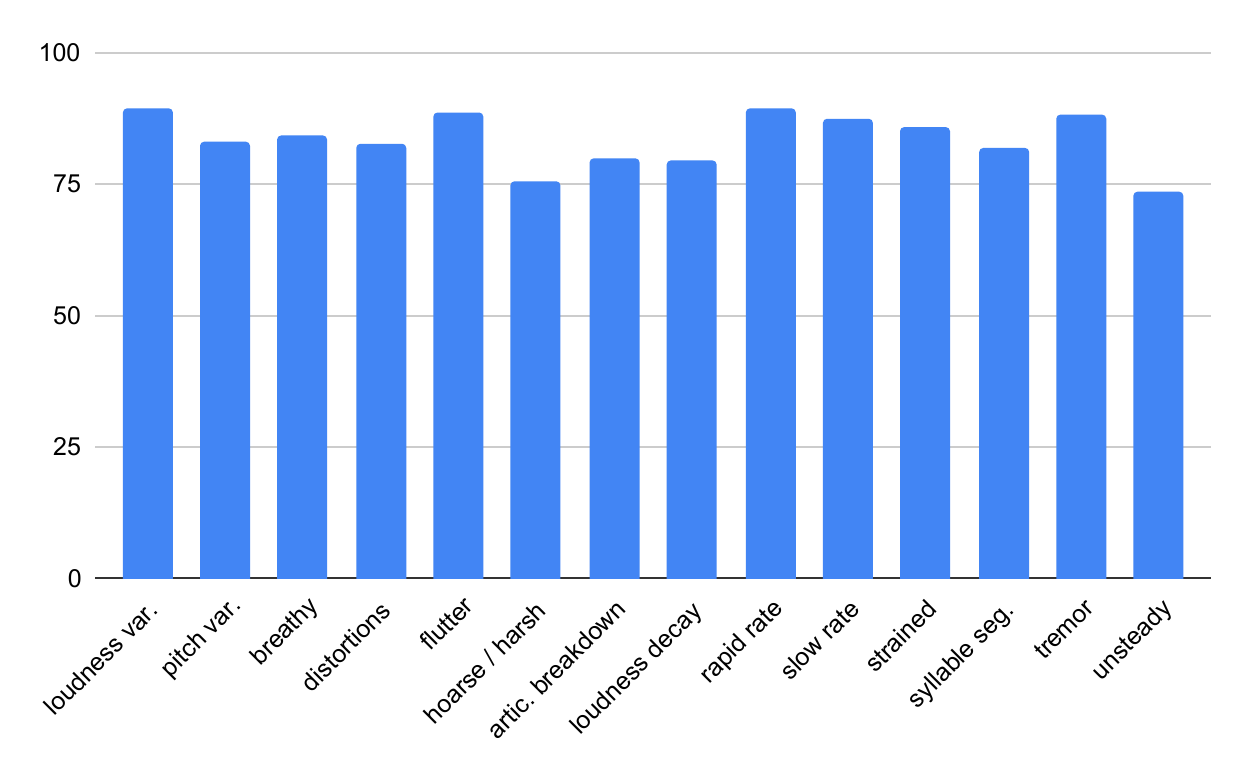}
 \caption{Prediction accuracy for each attribute (label) of the Medical Diagnostics Model using the Perceiver architecture.}
 \label{fig:full_results}
\end{figure}

%% file: conclusions.tex
\section{Conclusions}
\label{sec:conclusions}
In summary, we described a model that can predict attributes from speech that are important to clinicians when diagnosing neurological conditions. For this task, we proposed a factored perceiver-based architecture. The benefits of our model are two-folds. First, the model maps a sequence of audio encodings to a fixed set of class-specific latent representations, making it easier to build a classifier for the input sequence, when different classes require information from different regions of the signal. In contrast, prior approaches use a pooling layer prior to the logit layer, which may result in a loss of information in long sequences. Our factorized projection allows different classifiers to capture different time and feature correlations while also being data efficient. Our model obtains better prediction accuracy (83.1\%) over strong baselines using transformer (80.9\%) and conventional perceiver (81.8\%) models. We investigate the use of a (frozen) Universal Speech Model to provide audio encodings. USM provides large gains in prediction acccuracy. Furthermore, we show that USM trained for ASR works well even for tasks that are unrelated to ASR. Importantly, encodings from the middle layers provide a mix of both acoustic and phonetic information and achieve best prediction results (83.1\% vs 79.6\%). In future work, we will explore representations from different layers since our analysis suggests that modeling a diverse set of attributes such as ours can benefit encodings from different layers.